# Anomalous Nernst thermopower and giant magnetostriction in microwave synthesized $La_{0.5}Sr_{0.5}CoO_3$

Marimuthu Manikandan, Arup Ghosh and Ramanathan Mahendiran[1]

*Department of Physics, National University of Singapore, 2 Science Drive 3,*

*Singapore 117551, Republic of Singapore*

## Abstract

Ferromagnetic metallic oxides have potential applications in spincaloric devices which utilize the spin property of charge carriers for interconversion of heat and electricity through the spin Seebeck or the anomalous Nernst effect or both. In this work, we synthesized polycrystalline $La_{0.5}Sr_{0.5}CoO_3$ by microwave irradiation method and studied its transverse thermoelectric voltage (Nernst thermopower) and change in the linear dimension of the sample (Joule magnetostriction) in response to external magnetic fields. In addition, magnetization, temperature dependences of electrical resistivity and longitudinal Seebeck coefficient ($S_{xx}$) in absence of an external magnetic field were also measured. The sample is ferromagnetic with a Curie temperature of $T_C$ = 247 K and shows a metal like resistivity above and below $T_C$ with a negative sign of $S_{xx}$ suggesting charge transport due to electrons. Magnetic field dependence of the Nernst thermopower ($S_{xy}$) at a fixed temperature shows a rapid increase at low fields and a tendency to saturate at high fields as like the magnetization. Anomalous contribution to $S_{xy}$ was extracted from total $S_{xy}$ measured and it exhibits a maximum value of ~ 0.21 µV/K at 180 K for $H$ = 50 kOe, which is comparable to the value found in single crystal for a lower Sr content. The Joule magnetostriction is positive, i.e., the length of the sample expands along the

---

direction of the magnetic field and it does not saturate even at 50 kOe. The magnetostriction increases with decreasing temperature below $T_C$ and reaches a maximum value of 500 ppm at T ≤ 40 K. Coexistence of the anomalous Nernst thermopower and giant magnetostriction in a single compound has potential applications for thermal energy harvesting and low-temperature actuators, respectively.

# Introduction

For the past few years, a branch of spintronics known as Spin caloritronics is gaining much attention among researchers working in traditional thermoelectric or magnetic materials.[1] As the name implies, Spin caloritronics is an interdisciplinary field that bridges magnetism and heat transport.[2] While the spin degree of freedom of charge carriers is irrelevant for waste heat to energy conversion involving traditional thermoelectric materials such as $Bi_2Sb_3$, $PbTe_2$ etc., spin caloritronics aims to seek correlations between the heat, charge, and spin currents with a goal of harvesting electrical energy from waste heat using the spin property of electrons. Two spin caloritronic phenomena are considered to be attractive for energy conversion: The spin Seebeck effect (SSE) and the anomalous Nernst effect (ANE). The SSE occurs in a FM/NM bilayer film or a FM layer with NM strips attached on the top, where FM is a ferromagnetic metal[3] or a ferrimagnetic insulator (e.g., YIG)[4] and NM is a nonmagnetic heavy metal with a strong spin-orbit coupling such as Pt or Ta. In SSE, the FM layer experiences a thermal gradient that generates spin current and it is pumped into the NM layer/strip, where it is converted into a charge current through spin-orbit interaction in the NM layer, and finally detected as a dc voltage via inverse spin Hall effect. On the other hand, the ANE can be realized in a single layer magnetic film without a need for NM layer[5,6,7] or multilayer films[8] and in single crystals.[9]

The ANE is a transverse thermomagnetic effect. If a temperature gradient ($\nabla_x T$) is applied along *x*-axis and magnetic field (*H*) along *y*-axis to a conductor, a transverse electric field ($E_y$) develops along the z-axis, from which we extract the Nernst thermopower as $S_{xy} = E_y/\nabla_x T$. While $E_y \propto H$ in a paramagnet, a large non-linear component develops in low fields for a ferromagnet, which is known as the anomalous Nernst effect (ANE). In a ferromagnet, $E_y$ due to ANE is proportional to the magnetization.[6] Some magnetic materials with non-colinear spin structures also exhibit ANE even though the net magnetization is very small. The origin of ANE in these low-moment samples is a current topic of intense investigations and analysis of existing results points to the relevance of Berry curvature of a conduction band at the Fermi level.[10]

Cobalt-based oxides are an interesting class of materials for the ANE phenomenon because their magnetic, electrical, and thermoelectric properties are severely influenced by valence and spin states of the Co ions and by oxygen coordination as established in layered $NaCo_2O_4$,[11,12] Sr-substituted perovskite $LaCoO_3$,[13–18] and Bi-Sr-CoO misfit cobaltites.[19] While the longitudinal Seebeck effect (thermopower), in which a voltage is measured at the two ends of a sample along the direction of the temperature gradient, has been extensively studied in Co-based oxides, transverse Seebeck effect under a magnetic field, i.e., the anomalous Nernst effect has been rarely reported. The ANE was first reported in single crystals of $La_{1-x}Sr_xCoO_3$ ($x = 0.17, 0.25,$ and $0.3$) thirteen years ago,[20] but its existence in polycrystalline samples was realized only this year in $Pr_{0.5}Sr_{0.5}CoO_3$[21] and $La_{0.7}Sr_{0.3}CoO_3$[22]. Electrical conductivity and ferromagnetic Curie temperature ($T_C$) in $La_{1-x}Sr_xCoO_3$ series increase with increasing Sr content and $T_C$ reach a maximum ($T_C \sim 250$ K) for x = 0.5.[23–25] The compositions with $x > 0.5$ are prone to severe oxygen deficiency when synthesized under ambient air atmosphere but the end compound $SrCoO_3$ show ferromagnetic transition very close to room temperature ($T_C \sim 305$ K) when it fully oxidized.[26]

There is no report so far on the ANE for $x > 0.3$ in the $La_{1-x}Sr_xCoO_3$ series. Here, we report the ANE in ceramic $La_{0.5}Sr_{0.5}CoO_3$ for the first time along with a comprehensive study of magnetism, electrical resistivity, longitudinal thermopower and magnetostriction. We exploited microwave (MW) irradiation to synthesis the perovskite $La_{0.5}Sr_{0.5}CoO_3$. MW-synthesis is an emerging field for materials synthesis due to its great advantage of synthesis of the final product within few tens of seconds to tens of minutes compared to several days needed for solid-state synthesis using the electrical furnace.[27–30] While a sample in a conventional furnace is heated from outside to inside by convection of heat from heating elements to the sample, high-power microwave irradiation heats the sample internally. Reorientations of electrical dipoles in the sample absorb a maximum microwave power when they are driven to resonance by microwave electric field and the absorbed power is dissipated internally as volumetric heating. The temperature of the sample can increase as high as 1000 ºC to 1200 ºC in few tens of minutes which makes constituents of precursors react rapidly.[31] Besides electrical dipole reorientation, microwave power absorption also occurs due to eddy current, magnetic hysteresis and ferromagnetic resonance.[32] Interestingly, $CoFe_2O_4$ and some other oxides show an unusual recrystallization effect when exposed to the magnetic field component of microwave electromagnetic field.[33] Reports on MW synthesis of perovskite cobaltites are scarce. Recently, J. Guierrez Seijas *et al*. reported MW synthesis of undoped nanoparticles of $RCoO_3$ (R = La, Pr, etc) series.[34] Y.-F. Liu *et al*.[35] reported microwave synthesis of $La_{1-x}Sr_xCoO_3$ ($x = 0.10 - 0.50$) nanowires. In both the above works, electrical, magnetic or magneto thermoelectric properties were not reported.

**Experimental**

Stoichiometric proportions of $La_2O_3$, $SrCO_3$ and $Co_3O_4$ precursors were mixed well by grinding them in an agate mortar and pestle. $La_2O_3$ powder was dehydrated at 900 °C for 8 hr, to remove carbonates, before mixing. The mixed powder was transferred to alumina crucibles and it was placed in a MW furnace (Milestone PYRO microwave furnace, model MA 194-003). The powder was irradiated with a microwave frequency of 2.45 GHz at a power level of 1600 W in air atmosphere. The temperature was set to reach 1200 °C in 15 min by microwave irradiation and then maintained at 1200 °C for 20 min. Then, the microwave power was switched off and the pellet was allowed to cool to room temperature in 4 hours. The resultant fragments were ground into powder and characterized using X-ray diffraction (XRD) with Cu-K$\alpha_1$ (1.5406 Å) radiation for structural analysis. Morphological features and elemental identification were carried out using Field emission scanning electron microscopy (FE-SEM, JEOL JSM-6700F) with energy dispersive X-ray analysis (EDAX). Magnetic measurements were done using a vibrating sample magnetometer probe attached to a physical property measurements system (PPMS, Quantum Design, USA). The microwave synthesized powder was made as a thin rectangular pellet and sintered at 1000 °C for 3 hours in a box furnace for the thermopower and resistivity measurements. DC Electrical resistivity ($\rho$) and Seebeck coefficient ($S_{xx}$) were simultaneously measured at each stabilized temperature while cooling from 300 K to 50 K in a Janis closed-cycle refrigerator using a home-made four-probe setup where the thermal gradient was created by resistor heater mounted on the separated copper blocks and the sample bridges the block with thermally conductive grease.[36] A temperature difference of 2 to 3 K was maintained between two ends of the sample. Anomalous Nernst Effect (ANE) measurements were done up to a magnetic field of 50 kOe at selected temperatures in PPMS using a homely designed sample holder. A polished rectangular sample in the dimension of ~ 5 × 4 × 0.5 mm$^3$ was bridged on two separated copper blocks which were electrically insulated from the sample using a 15 μm thick kapton tape. The thermal gradient was produced in the sample by two cartridge heaters installed in holes drilled in the copper blocks. The sample temperatures were regulated by heaters and measured by two calibrated Cernox thermometers mounted on the surface of each copper blocks. The temperature of the sample was taken from the mean value of temperatures at hot and cold copper blocks. A temperature difference of 10 K between hot and cold ends was maintained, which corresponds to a temperature gradient of 2 K/mm. The thermal gradient, applied magnetic field and measured voltage were kept perpendicular to each other. ANE voltages were measured using a Keithley 2182A nanovoltmeter and calculated as $V_{ANE} = (V(+H) - V(-H))/2$ to eliminate the contribution from background signal and longitudinal magneto-thermopower. The temperature and field dependence of linear magnetostrictions were measured along the magnetic field direction (longitudinal) in the temperature range of 300 K - 10 K using a capacitance dilatometer probe designed for PPMS. The field sweep was maintained at 50 Oe/sec for both magnetization and magnetostriction measurements.

**Results and discussion**

**Structural and morphological studies**

The powder X-ray diffraction pattern of microwave synthesized $La_{0.5}Sr_{0.5}CoO_3$ is shown in Fig. 1. The sample is found to be single-phase and possess a rhombohedral structure with space group $R\bar{3}c$. The refined cell parameters at room temperature are **a** = **b** = 5.416(2), **c** = 13.266(5) Å and volume **V** = 337 Å$^3$ which are close to the parameters reported for $La_{0.5}Sr_{0.5}CoO_3$ synthesized at oxygen atmosphere with slow cooling.[37] $LaCoO_3$ has a rhombohedral structure and the rhombohedral distortion in $La_{1-x}Sr_xCoO_3$ decreases as Sr content increases, and a cubic phase emerge for $x > 0.5$.[38–40] Sunstrom et al.,[41] reported a gradual change in the structure with increasing $x$ in chemically oxidized perovskite samples of $La_{1-x}Sr_xCoO_{3-\delta}$ ($0.5 \leq x \leq 0.9$), as indicated by splitting/broadening of the diffraction peaks. The diffraction pattern of microwave synthesized $La_{0.5}Sr_{0.5}CoO_3$ also shows a splitting/doublet in their peaks but they are noticeable only at higher Bragg angles rather than at lower angles, shown in the inset of Fig. 1, which may be due to the pseudo-cubic nature of the sample. Similar kind of local structural inhomogeneity was also reported by Haggerty et al.,[37] and Karpinsky et al.,[42] in solid-state synthesised $La_{0.5}Sr_{0.5}CoO_{3-\delta}$. Hence, short time microwave synthesised material seems to be identical to the material synthesised by a conventional solid-state method.

Field emission scanning electron micrographs of as synthesised and post hardened (1000 °C for 12 h) samples are shown in Fig. 2(a) & 2(b), respectively. An irregular shaped flaky natured particles with heterogeneous size distribution are realised in both as synthesised and post hardened samples. The particles are fragmented between micron and nanometer ranges.

The largest is about 3 microns and the smallest is ~ 100 nanometers. In microwave heating, heat creates inside the material initially and then transfers to the surface, which results in tiny broken fragments dappled on the surface of the particles, whereas the post sintered sample shows a smooth surface due to the gradual heating and cooling process. The magnified surface images of both the samples are shown as insets in the respective images. The sole existence of La, Sr, Co and O elements in the EDAX spectrum, Fig. 2(c), confirms the purity of the synthesized sample.

**Magnetization, electrical transport and thermopower**

Fig. 3(a) shows the temperature dependence of magnetization ($M$) under $H$ = 1 kOe magnetic field measured during cooling from 350 K to 10 K. The rapid increase of $M(T)$ around 250 K signals a phase transition from paramagnetic to a ferromagnetic state while cooling. From the minimum of d$M$/d$T$ curve, shown in the same figure, we obtain a ferromagnetic Curie temperature of $T_C$ = 247 K which is close to the $T_C$ reported by other researchers for stoichiometric $La_{0.5}Sr_{0.5}CoO_3$ synthesized by the standard solid-state reaction method.[37–39,42]

Fig. 3(b) shows the $M$-$H$ isotherm at 10 K and the inset shows hysteresis in an enlarged scale. The coercive field is 530 Oe at 10 K. $M$ shows a tendency to saturate above 40 kOe and the maximum value of $M$ is ~ 1.89 $\mu_B$/f.u. at 50 kOe, which is close to the magnetization values (~ 1.8-1.9$\mu_B$/f.u.) reported for the solid-state synthesized stoichiometric $La_{0.5}Sr_{0.5}CoO_3$.[39,42,43] By considering the combinations of the spin only contributions from intermediate spin states of $Co^{3+}$ (S= 1) and $Co^{4+}$ (S = 2), the saturation magnetization is expected to be $M$ = 2.5 $\mu_B$/Co-ion. However, a strong hybridization of Co-3$d$ and O-2$p$ orbitals can result in a lower value of saturation magnetization because of itinerant electron like band structure.[44]

Fig. 4 shows the temperature dependence of resistivity ($\rho$) and thermopower ($S$) in the temperature range $T$ = 350 K to 50 K measured in a zero magnetic field. $\rho(T)$ exhibits metallic behavior in the entire temperature range, but with a change of slope around $T_C$ as a consequence of a decrease in spin-disorder scattering of charge carriers. The sign of $S$ is negative in the measured temperature range and it suggests that the majority charge carriers here are electrons. While $S$ decrease nearly linearly as $T_C$ is approached, the change is rapid just below $T_C$. Then, $S(T)$ shows an upturn around 150 K and an increase (the magnitude decreases towards zero) on further cooling, which is expected since $S$ being the measure of entropy as to go to zero at absolute temperature.

**Anomalous Nernst effect**

Fig. 5 shows the Nernst thermopower ($S_{xy}$) as a function of magnetic field at fixed temperatures for the temperature range (a) $T = 300$ K – 200 K and (b) $T = 180$ K to 25 K. At 300 K, $S_{xy}$ increases linearly with increasing magnetic field as expected in a paramagnetic. When $T = 280$ K and 260 K, although $S_{xy}$ increases linearly with $H$ above 20 kOe, it increases non-linearly in lower fields. The slop of $S_{xy}$ at low field increases as temperature decreases below 240 K and the high field data nearly saturates below 100 K. The magnitude of $S_{xy}$ at 50 kOe enhances as the temperature lowers below 300 K, goes through a maximum value around 180 K and decreases at lower temperatures. The inset of Fig. 5(b) compares the field dependence of $S_{xy}$ (left side) and $M$ (right side) at 25 K. We can see a close resemblance: both show a rapid increase in low fields ($H < 4$ kOe) and a tendency to saturate at high fields ($H \geq 40$ kOe). The measured signal in the ferromagnetic regime $S_{xy} = S_{xy}(NE) + S_{xy}(ANE)$ is a sum of the normal ($S_{xy}(NE)$) and the anomalous ($S_{xy}(ANE)$) contributions. Since $S_{xy}(NE) \propto H$, the ANE part is extracted from the measured $S_{xy}$ by subtracting linear extrapolation of the high-field portion of $S_{xy}$ vs $H$ curves to zero field. The extracted $S_{xy}(ANE)$ for $H = 5$ kOe and 50 kOe are plotted against the temperature in Fig. 6. Starting from low temperature, $S_{xy}(ANE)$ increases with increasing temperature and shows a maximum of ~ 0.21 μV/K for 50 kOe around 180 K afterwards it decreases rapidly and vanishes above $T_C$. The maximum value of $S_{xy}$ ~ 0.21 µV/K is lower than the value observed in single crystal (0.26 µV/K)[20] and little higher than the polycrystalline (~ 0.19 µV/K)[21] of $La_{0.7}Sr_{0.3}CoO_3$. The temperature dependence of $S_{xy}$ well below $T_C$ was fitted for $H = 50$ kOe using Eq. (1)[45,9] and shown in the inset of Fig. 6, where the solid line represents the fitted curve. This equation relates the $S_{xy}(ANE)$ to the longitudinal resistivity ($\rho_{xx}$) and Seebeck coefficient ($S_{xx}$).

$$S_{xy}(ANE) = \rho_{xx}^{(n-1)} \left[ \frac{\pi^2 k_B^2}{3e} \lambda' T - (n-1)\lambda S_{xx} \right] \quad (1)$$

Here, $\lambda'$ is the first-order energy derivative of $\lambda$ and $n$ is a positive number that connects the Hall resistivity, $\rho_{xy}$ and linear resistivity, $\rho_{xx}$ through $\rho_{xy} = \lambda \rho_{xx}^n$. For intrinsic (Berry curvature related) and extrinsic (side-jump scattering) mechanisms, $n = 2$. However, if the extrinsic skew-scattering mechanism dominates, $n$ becomes 1 for a clean metal. In case of a bad metallic conduction regime, $n \sim 0.4$ where the electrical conduction takes place due to hopping of the charge carriers.[20] The fitted parameters are $\lambda \sim -1 \times 10^{-4}$, $\lambda' \sim 2 \times 10^{15}$ and $n = 0.55 \pm 0.005$ for $H = 50$ kOe. The obtained value of $n$ indicates that the microwave synthesised $La_{0.5}Sr_{0.5}CoO_3$ is in a bad-metallic conduction regime where the major contribution to ANE is driven by the hopping of charge carriers. The occurrence of a broad maximum below $T_C$ can be qualitatively understood using a modified form of Eq.1: $S_{xy} = \frac{\rho_{xy}}{\rho_{xx}} \left[ \frac{\pi^2 k_B^2}{3e} T \frac{\lambda'}{\lambda} - (n-1)S_{xx} \right]$[45] where, $-\frac{\rho_{xy}}{\rho_{xx}} = tan\theta_H$ (Hall angle). The decrease of $S_{xy}$ much below $T_C$ is not pertinent to polycrystalline nature of the sample but also found in single crystals.[20,22] The $S_{xy}$ is strongly influenced by the magnetic field dependences of the Hall angle and $S_{xx}$. The $S_{xy}$ increases rapidly just below $T_C$ because of the increase of magnetization. When the magnetization saturates much below $T_C$, the decrease in $S_{xx}$ and the Hall angle with temperature dominates and hence $S_{xy}$ also decreases.

## Magnetostriction

Magnetostriction refers to a change in the physical dimensions of a sample when it is magnetized. The change is mostly in length (Joule magnetostriction) with the preservation of volume in most materials much below $T_C$. Fig. 7(a) shows the field dependence of Joule (parallel) magnetostriction, $\lambda_{par} = [L(H) - L(0)]/L(0)$ where $L(H)$ is the length of the sample in a magnetic field $H$ measured along the direction of the applied magnetic field, and magnetization at 10 K for comparison. $\lambda_{par}$ increases gradually and nonlinearly as the magnetic field increased from 0 kOe to 50 kOe without showing a saturation. While decreasing the field $\lambda_{par}$ retains higher values in the field range $H \sim 15 - 50$ kOe, but overlaps with the field increasing part of the curve in lower fields. In Fig. 7(b), we show the hysteresis loops only in the first quadrant for different temperatures, from 20 K to 280 K. The width of hysteresis narrows gradually with increasing temperature and vanishes above $T_C$. We extract the value of $\lambda_{par}$ at the maximum field ($H = 50$ kOe) from the isothermal field sweep data and plot them as a function of temperature in Fig. 7(c). $\lambda_{par}$ is negligibly small in the paramagnetic state but it increases with lowering the temperature below $T_C$. Below 40 K, $\lambda_{par}$ is nearly temperature independent and reaches a maximum value of $\sim 500 \times 10^{-6}$ or 500 ppm (parts per million). This value is larger than the value (400 ppm at 25 K for $H = 50$ kOe) reported for a solid-state synthesized $La_{0.5}Sr_{0.5}CoO_3$ sample using a pulsed magnetic field.[46] In this, the magnetostriction was measured with the strain gauge method, and the striction was recorded only while decreasing the magnetic field from a maximum value of $H = 142$ kOe. In the pulsed experiment, magnetostriction parallel and perpendicular to the applied magnetic field were measured and they were found to have opposite signs and different values. However, we cannot measure magnetostriction perpendicular to the magnetic field with the existing capacitance dilatometer.

Reports on the magnetostriction in perovskite oxides are scarce even now.[46–48] Magnetostriction in a material is inherently connected to spin-orbit interaction. Since the orbital moment of transition metal ions such as $Fe^{3+}$, $Co^{3+}$, and $Ni^{3+}$ in octahedral crystalline electric field is generally quenched, magnetostriction is expected to be small in perovskite oxides containing these transition metal ions. It was proposed that both hole-rich ferromagnetic and hole-poor non-magnetic phases coexist within a single crystallographic structure of $La_{0.5}Sr_{0.5}CoO_3$ sample synthesized by conventional solid-state reaction method. In the non-magnetic phase, $Co^{3+}$ ions are in the low-spin state (S = 0) which is orbitally non-degenerate. It was suggested that the applied magnetic field gradually transforms the low-spin $Co^{3+}$ ($t_{2g}^6$) ions into intermediate spin $Co^{3+}$ ($t_{2g}^5 e_g^1$) ions and create orbital singlet ($d_{xy}$) with a no-orbital moment and orbital doublet ($d_{yz}$, $d_{xz}$) with the unquenched orbital moment. Hence, spin-orbit coupling is induced by the magnetic field. When the spin rotates towards the field direction, $t_{2g}$-orbital also tend to rotate. As the $3d$-orbital couple to the lattice via crystal field interaction, lattice expands along the field direction. Very recently, Subias et al.[49] carried out differential X-ray absorption with and without external magnetic field in $La_{0.5}Sr_{0.5}CoO_3$ and $CoFe_2O_4$, and concluded that the Co-O bond length along [100] axis elongates under a magnetic field in $La_{0.5}Sr_{0.5}CoO_3$ but shrinks in $CoFe_2O_4$. The observed hysteresis in magnetostriction in the high-field region, above technical saturation, suggests a slow relaxation of lattice while decreasing the field and it does not impact the magnetization. There are two possibilities: 1. There is a structural transition from rhombohedral structure at low fields ($H$ < 10 kOe) to cubic at high fields ($H$ > 10 kOe) and the hysteresis at high fields is due to the coexistence of these two structures. The structure reverses back to rhombohedral when the field is reduced to zero, 2. For the fields above 10 kOe, twin deformation (rearrangement of structural domains) occurs. Very recently, Yokosuko et al.[50] observed twin deformation in single-crystalline of $La_{0.8}Sr_{0.2}CoO_3$ even in the paramagnetic phase. Magnetostriction in the initial field sweep (0

→ 90 kOe) was different from the subsequent field sweeps, which showed hysteresis over a wide field range in their sample. We need to verify such a possibility in our sample through X-ray or microscopy under external magnetic fields.

## Conclusion

We have successfully synthesized a single-phase $La_{0.5}Sr_{0.5}CoO_{3-\delta}$ by microwave irradiation method. Field dependence of the Nernst thermopower and magnetostriction were measured. The anomalous Nernst effect shows a rapid increase around $T_C$ (247 K) and exhibits a maximum value of ~ 0.21 µV/K at 180 K in an applied magnetic field of 50 kOe. The sign of magnetostriction is positive and its magnitude below $T_C$ smoothly increases with decreasing the temperature down to 40 K below which it is nearly constant. The fractional change in the length of the sample is 500 ppm at 40 K and at 50 kOe, which is higher than the previously reported value in the solid-state synthesized sample. Also, the possibility of field-induced structural transition in this sample needs to be investigated. Since microwave sintering is a fast and energy-efficient technology, it will be interesting for future work to investigate how the values of the anomalous Nernst effect and Joule magnetostriction are influenced by grain size, morphology, and density of the sample.

**Conflicts of interest**

There are no conflicts to declare.

**Acknowledgements**

R. M. acknowledges the Ministry of Education, Singapore, for supporting this work (Grants numbers: R1444-000-404-114 and R144-000-422-114).

**Figure captions**

**Fig. 1** X-ray diffraction pattern of microwave synthesized $La_{0.5}Sr_{0.5}CoO_3$. Inset shows a magnified view of the peak splitting in higher and lower angles.

**Fig. 2** FE-SEM images of **(a)** as microwave irradiated, **(b)** post sintered $La_{0.5}Sr_{0.5}CoO_3$ and **(c)** EDAX spectrum of as microwave irradiated sample. Insets show an enlarged view of particle surfaces.

**Fig. 3 (a)** Temperature-dependent magnetization (*M*) of $La_{0.5}Sr_{0.5}CoO_3$ showing ferromagnetic transition at $T = T_c$ under $H = 1$ kOe. The right-y axis shows d*M*/d*T*. **(b)** M-H isotherm at 10 K. Inset shows the enlarged view of hysteresis.

**Fig. 4** Temperature-dependent resistivity ($\rho$) and Seebeck coefficient (S) of the microwave synthesized $La_{0.5}Sr_{0.5}CoO_3$ in the temperature range 300 K - 50 K. FM is ferromagnetic region and PM is paramagnetic region.

**Fig. 5** Magnetic field dependence of the Nernst thermopower ($S_{xy}$) of the microwave synthesized $La_{0.5}Sr_{0.5}CoO_3$ measured at various temperatures **(a)** $T = 300$ K – 200 K and **(b)** $T = 180$ K – 25 K. Inset shows a comparison of $S_{xy}$ (left y-axis) and *M* (right-y-axis) versus *H* at 25 K.

**Fig. 6** Temperature dependence of the anomalous Nernst thermopower $S_{xy}$ (ANE) in microwave synthesized $La_{0.5}Sr_{0.5}CoO_3$ extracted for different strengths of the magnetic field ($H = 5$ kOe and 50 kOe). Inset shows the $S_{xy}$ data and fit to Eq. (1) in the low-temperature region $T \sim 150$ K to 10 K. FM is ferromagnetic region and PM is paramagnetic region. Lines are guides to the eyes.

**Fig. 7 (a)** Hysteresis loop of Joule magnetostriction ($\lambda_{par}$) measured parallel to the direction of applied magnetic field and magnetization at 10 K **(b)** Joule magnetostriction ($\lambda_{par}$) isotherms at different temperatures **(c)** Temperature dependence of the $\lambda_{par}$ at the maximum magnetic field of $H = 50$ kOe. FM is ferromagnetic region and PM is paramagnetic region.

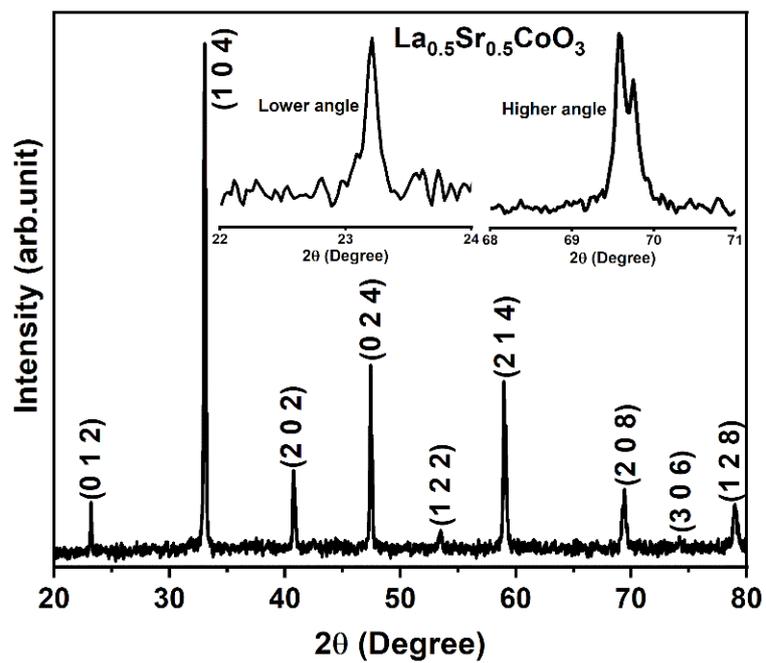

Fig. 1 Manikandan et al.

**Fig. 1** X-ray diffraction pattern of microwave synthesized $La_{0.5}Sr_{0.5}CoO_3$. Inset shows a magnified view of the peak splitting in higher and lower angles.

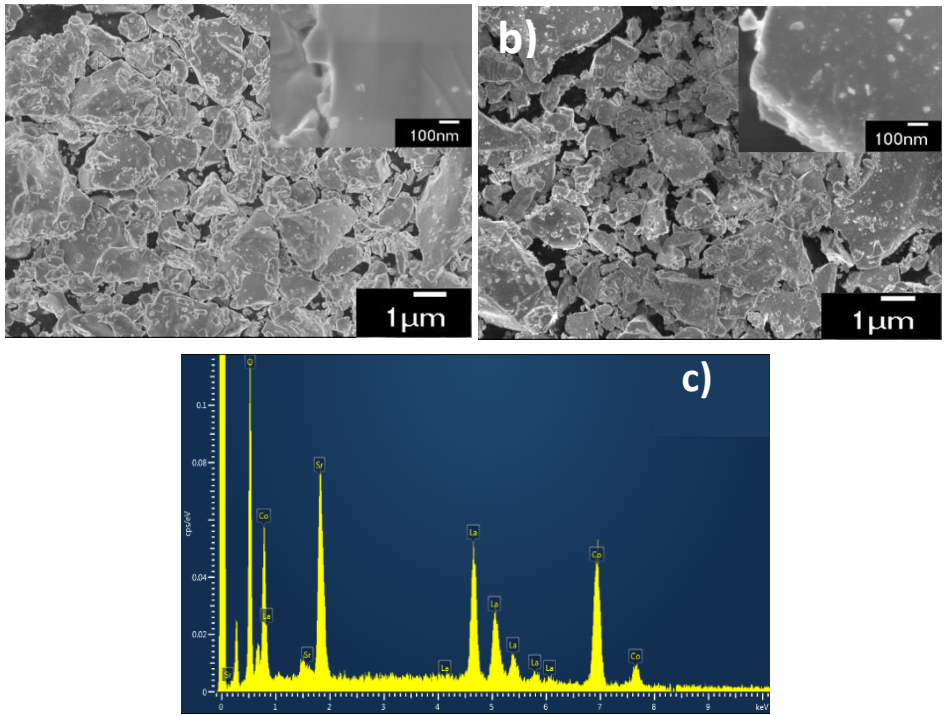

Fig. 2 Manikandan et al.

**Fig. 2** FE-SEM images of **(a)** as microwave irradiated, **(b)** post sintered $La_{0.5}Sr_{0.5}CoO_3$ and **(c)** EDAX spectrum of as microwave irradiated sample. Insets show enlarged view of particle surfaces.

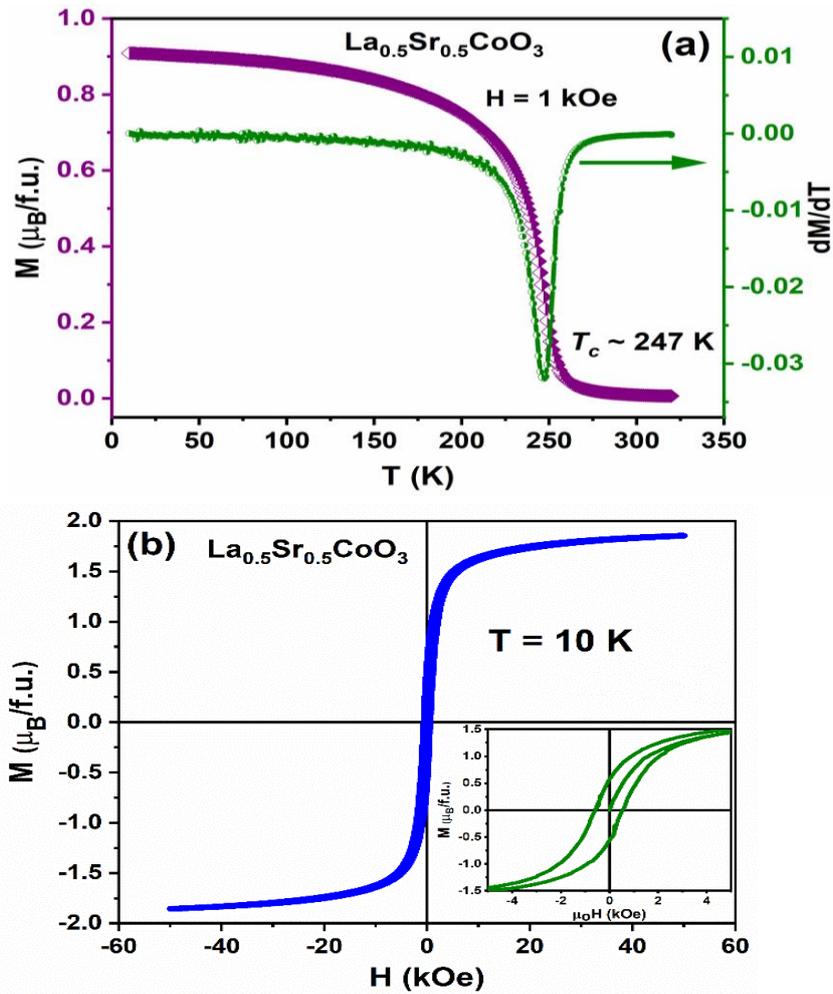

Fig. 3 Manikandan et al.

**Fig. 3 (a)** Temperature dependent magnetization (*M*) of La$_{0.5}$Sr$_{0.5}$CoO$_3$ showing ferromagnetic transition at $T = T_c$ under $H$ =1 kOe. The right-y axis shows d*M*/d*T*. **(b)** M-H isotherm at 10 K. Inset shows the enlarged view of hysteresis.

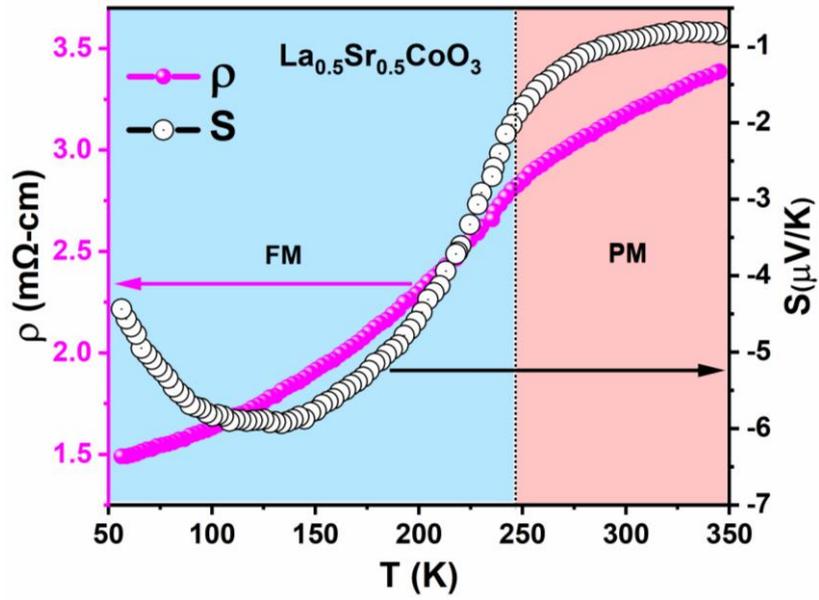

Fig. 4 Manikandan et al

**Fig. 4** Temperature dependent resistivity (ρ) and Seebeck coefficient (S) of the microwave synthesized $La_{0.5}Sr_{0.5}CoO_3$ in the temperature range 300 K - 50 K. FM is ferromagnetic region and PM is paramagnetic region.

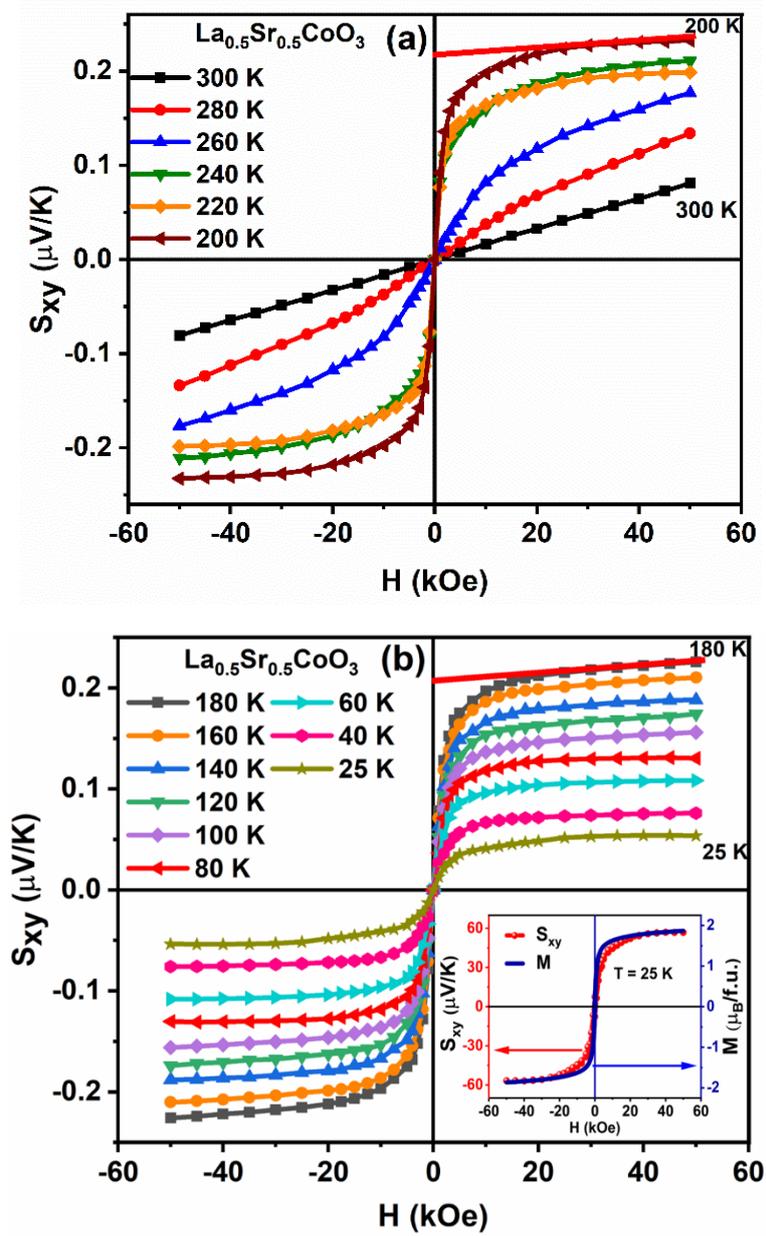

Fig. 5 Manikandan et al.

**Fig. 5** Magnetic field dependence of the Nernst themopower ($S_{xy}$) of the microwave synthesized La$_{0.5}$Sr$_{0.5}$CoO$_3$ measured at various temperatures **(a)** $T$ = 300 K –200 K and **(b)** $T$ = 180 K – 25 K. Inset shows comparison of $S_{xy}$ (left y-axis) and $M$ (right-y-axis) versus $H$ at 25 K.

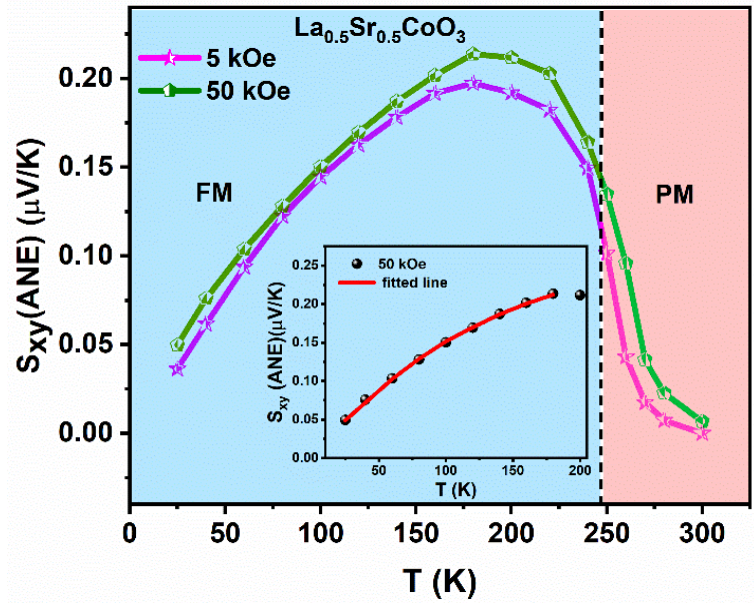

**Fig. 6** Temperature dependence of the anomalous Nernst thermopower $S_{xy}$ (ANE) in microwave synthesized $La_{0.5}Sr_{0.5}CoO_3$ extracted for different strengths of the magnetic field ($H$ = 5 kOe and 50 kOe). Inset shows the $S_{xy}$ data and fit to Eq. (1) in the low-temperature region $T$ ~ 200 K to 10 K. FM is ferromagnetic region and PM is paramagnetic region. Lines are guide to eyes.

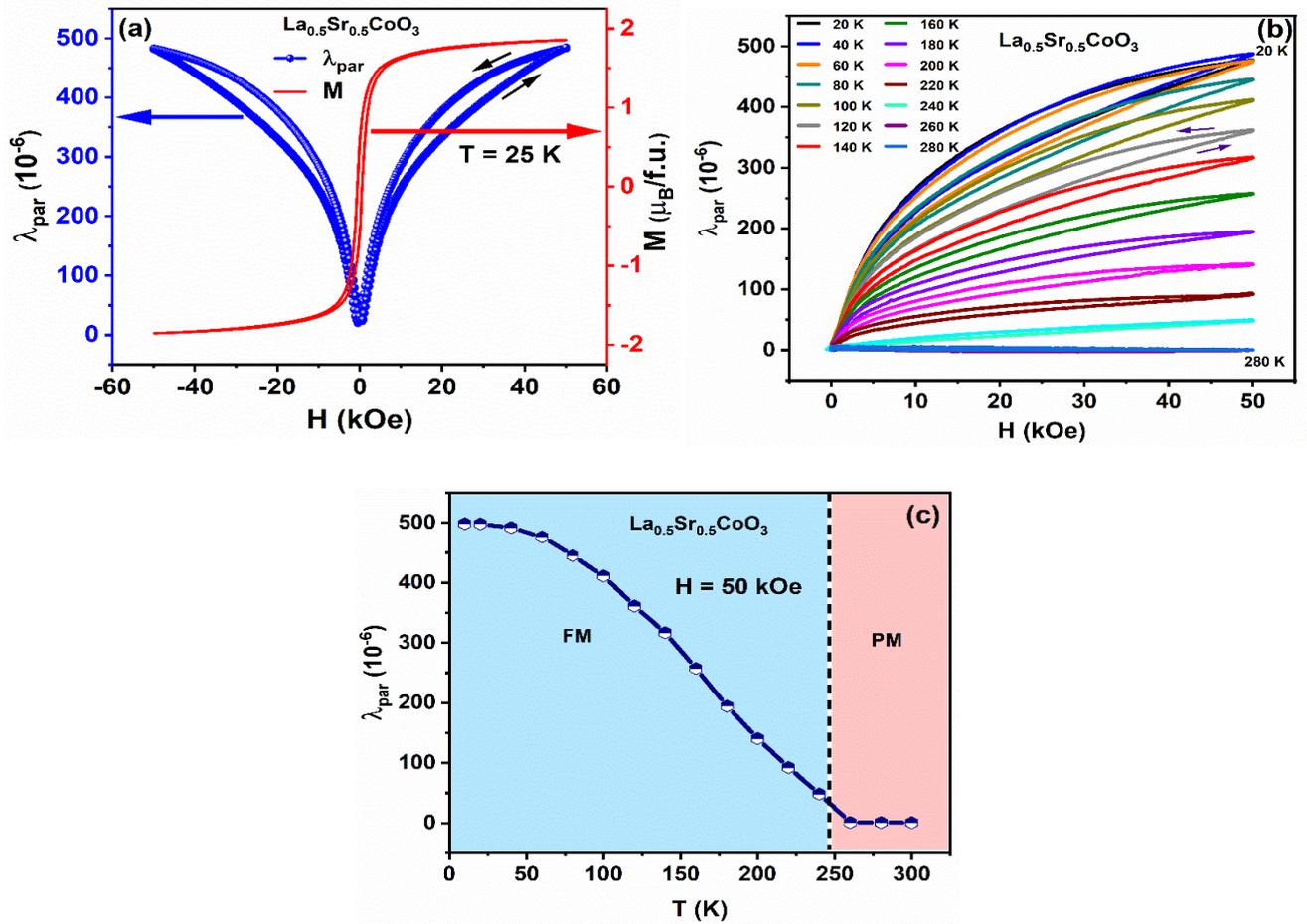

Fig. 7. Manikandan et al.

**Fig. 7 (a)** Hysteresis loop of Joule magnetostriction ($\lambda_{par}$) measured parallel to the direction of applied magnetic field and magnetization at 10 K **(b)** Joule magnetostriction ($\lambda_{par}$) isotherms at different temperatures **(c)** Temperature dependence of the $\lambda_{par}$ at the maximum magnetic field of $H$ = 50 kOe. FM is ferromagnetic region and PM is paramagnetic region.

23